# Bonding and oxidation protection of $Ti_2AlC$ and $Cr_2AlC$ for a Ni-based Superalloy

Maxim Sokol[1], Jian Wang[1], Hrishikesh Keshavan[2] and Michel W. Barsoum[1,*]

[1]*Department of Materials Science & Engineering, Drexel University, Philadelphia, PA 19104, USA*

[2] *GE Global Research, 1 Research Circle, Niskayuna NY 12309, USA*

[*]Corresponding author: Email – barsoumw@drexel.edu

**Abstract:**

Alumina forming, oxidation and thermal shock resistant MAX phases are of a high interest for high temperature applications. Herein we report, on bonding and resulting interactions between a Ni-based superalloy, NSA, and two alumina forming MAX phases. The diffusion couples $Cr_2AlC$/Inconel-718/$Ti_2AlC$ were assembled and heated to 1000 or 1100°C in a vacuum hot press under loads corresponding to stresses of either 2 MPa or 20 MPa. The resulting interfaces were examined using X-ray diffraction, scanning electron microscopy and energy-dispersive X-ray spectroscopy. Good bonding between $Cr_2AlC$ and NSA was achieved after hot pressing at 1000°C and a contact pressure of only 2 MPa; in the case of $Ti_2AlC$ a higher temperature (1100°C) and pressure (20 MPa) were needed. In both cases, a diffusion bond was realized with no evidence of interfacial damage or cracking after cooling to room temperature. Twenty thermal cycles from room temperature to 1000°C showed that $Ti_2AlC$ is a poor oxidation barrier for Inconel-718. However, in the case of $Cr_2AlC$ no cracks, delamination nor surface degradation were observed, suggesting that this material could be used to protect Inconel-718 from oxidation.

# 1. Introduction:

Nickel-base superalloys, NSAs, were designed to be used in the hottest parts of a gas turbine.[1] They exhibit a combination of mechanical strength and resistance to surface degradation unmatched by any other metallic alloys and/or compounds.[2] Due to those unique combination of properties, NSAs were also successfully applied in the oil and gas industry, space vehicles, submarines, nuclear reactors, military electric motors, chemical processing vessels and heat exchanger tubing.[2,3] Modern turbine engines rely heavily on NSAs for high temperature components and constitute over 50 % of the weight of advanced aircraft engines.

An increase in the operating temperatures of a turbine engine will result in improved performance and efficiency.[4] With a constant demand for increased operating temperature of gas turbine engines, it would be beneficial to develop robust and advanced protective materials. Currently, NSA protective coatings are generally based on advanced thermal barrier coatings, TBCs, that consists of two, or more, layers: a ceramic top coating, TC, and an underlying metallic bond coating, BC.[5,6] Typically, MCrAlY, where M is Ni and/or Co and NiAl-based alloys are used as BCs in TBCs. Enhanced adhesion of MCrAlY, as a TC layer, to the substrate would provide improved oxidation and corrosion protection to NSAs.[7] In both MCrAlY and NiAl-based alloys, the Al rich β phase serves as an Al source for the formation of continuous and protective $Al_2O_3$ scales. However, after prolonged exposure to high temperature, Al depletion of the BC leads to Al diffusion from the NSA substrate[8,9] and to the formation of harmful secondary reaction zones, that can ultimately deteriorate the mechanical properties of the coated substrate and decreases the TBCs' overall durability and lifetime.[10]

The MAX or $M_{n+1}AX_n$ phases are layered, hexagonal, early transition-metal carbides and nitrides, where n = 1, 2, or 3, ''*M*'' is an early transition metal, ''*A*'' is an A-group (mostly groups 13 and 14) element, and ''*X*'' is C and/or N.[11–13] Those materials have attracted increasing attention as candidates for new generation BCs[14–17] as some of them combine some of the more attractive properties of metals (machinability,[18] high thermal and electric conductivities,[11,12,19] excellent thermal shock resistance and damage tolerance)[18] and ceramics (high melting points, lightweight, high stiffness values and creep,[20–23] oxidation,[24–26] and fatigue[27] resistant). Alumina forming MAX phases, namely $Ti_3AlC_2$, $Ti_2AlC$, and $Cr_2AlC$ are also quite oxidation resistant (up to 1400°C) because they form slow growing protective alumina, $Al_2O_3$, layers [26,28–33]. Furthermore, studies based on growth models, documented that the grain boundary diffusion, in the $Al_2O_3$, layers scales are in accord with those obtained for one of the best oxidation resistant, alumina-forming FeCrAl alloys.[34,35]

Another beneficial characteristic of some of the MAX phases is its coefficient of thermal expansion (CTE),[13] which in the case of $Ti_2AlC$ ($8.2\times10^{-6}$/°C)[36,37] is a better match to α-$Al_2O_3$ ($9.3\times10^{-6}$/°C),[38] as compared directly to NSA (15–16$\times10^{-6}$/°C),[39] ensuring better adherence.[30,31] In the case of $Cr_2AlC$, the CTE value (12-13$\times10^{-6}$/°C)[36,37] lies between the CTE values of α-$Al_2O_3$ and NSAs that again prevents spallation, caused by thermal stresses.[40]

Smialek at el. studied the high temperature oxidation of the best known alumina forming MAX phases, $Ti_2AlC$ and $Cr_2AlC$, under realistic turbine engine environments and coating configurations.[14-17] The Cr$_2$AlC phase showed relatively good mechanical and chemical stability with a low γ' solvus and high refractory content Ni-based superalloy, LSHR, after being exposed to air at 800°C for up to 1000 h. It follows that Cr$_2$AlC is a promising candidate for high temperature, strain tolerant, corrosion resistant coating.[14] The same group plasma sprayed an yttrium-stabilized zirconia, YSZ, TBC on Ti$_2$AlC substrates. Oxidation experiments, performed at temperatures between 1100 to 1300°C, showed an extreme oxidative TBC durability due to a good match between the CTEs of Ti$_2$AlC and α-Al$_2$O$_3$.[15]

In a recent study, Gonzales-Julian et al., thermally cycled Cr$_2$AlC between 250 and 1200 °C using a burner rig. After an accumulated time of 29 h at the maximal temperature no cracks, delaminations or other damage was observed.[41] In another study, by the same group, TBCs based on YSZ were deposited, using atmospheric plasma spray, on dense Cr$_2$AlC substrates and thermally cycled at temperatures between 1100 and 1300°C.[42] Due to the formation of an α-Al$_2$O$_3$ layer the composite survived for 500 h at 1100°C without showing any damage. However, after 500 h at 1200°C, some cracks were observed. Finally, at 1300°C the system catastrophically failed after 268 h due to the formation of porous carbide layer underneath the α-Al$_2$O$_3$ layer.

The purpose of the present work is to examine the compatibility of Cr$_2$AlC and Ti$_2$AlC on Inconel 718. The reactivity between MAX/NAS, effect of bonding conditions and additional thermal cycling to 1000°C in air were investigated. To the best of our knowledge, this is the first report on the interaction of Inconel 718 with any MAX phases.

## 2. Materials and Experimental Details.

The $Cr_2AlC$ powders were prepared from a 2:1:1.05 stoichiometric mixture of Cr (-325 mesh, 99%, Alfa Aesar, Ward Hill, MA), Al (-325 mesh, 99.5%, Alfa Aesar, Ward Hill, MA), and graphite (-325 mesh, 99%, Alfa Aesar, Ward Hill, MA) powders. For $Ti_2AlC$, TiC (2–3 µm average particle size, 99.5% purity, 99%, Alfa Aesar, Ward Hill, MA), Ti, (-325 mesh, 99.5%, Alfa Aesar, Ward Hill, MA) and Al (-325 mesh, 99.5%, Alfa Aesar, Ward Hill, MA) powders were mixed in 0.95:1.05:1.1 molar ratio. Both mixtures were ball milled for 24 h in a polyethylene jar using zirconia milling balls with 1:2 weight ratio.

The powders were loaded into a graphite die, coated with boron nitride and densified in a hot press, HP, with graphite heating elements and vacuum of < 10 Pa. The sintering conditions were as follow: peak temperature 1400°C, heating rate of 7.5 °C/min, uniaxial pressure of about 36 MPa and dwell time of 4 h. The dense HPed surfaces were ground with coarse SiC paper to remove residual graphite and boron nitride. Density was measured by Archimedes technique and phase composition by X-ray diffraction, XRD, using a Rigaku SmartLab diffractometer (Tokyo, Japan) over the range of 10 to 80 2θ degree, with a step size of 0.02 degree and a dwell time of 0.5 s per step.

The MAX samples were machined into 1-2 mm thick cylinders with diameters of 25 mm. Tablets with same diameter and a thickness of about 3 mm were sectioned from the Inconel 718 using electric discharge machining (EDM).

All the samples were later mirror polished up to 3 µm finish and ultrasonically cleaned in ethanol. The NSA tablets were placed in between the two different MAX samples and treated for 2 h at various temperatures and pressures: 1000°C under a load corresponding to a pressure of 2 MPa, 1100°C under 2 MPa and 1100°C under 20 MPa. The vacuum level in all cases was < 1 Pa. Upon cooling, the samples were cross-sectioned, mounted, ground, polished and etched for 2-3 s by a 1:1:1-part solution of water, nitric acid (68 %, Alfa Aesar, Ward Hill, MA) and hydrofluoric acid (48 %, Sigma Aldrich, St. Louis, MO).

The diffusion couples' interface microstructures and compositions were analyzed by a scanning electron microscope, SEM (FEI Zeiss Supra 50VP, Jena, Germany), equipped with an energy-dispersive X-ray spectroscope, EDS (Oxford INCA X-Sight 7573, Abingdon, England).

The diffusion couples were then placed in a box furnace and heated to 1000 °C at a rate of 10 °C/min and dwell time of 0.5 h, at which time they were removed from the furnace and allowed to cool in ambient air. This process was repeated 20 times.

## 3. Results and Discussion

### 3.1 Sample characteristics

The HPed $Cr_2AlC$ and $Ti_2AlC$ samples were fully dense with densities of 5.2 and 4.1 $g/cm^3$ respectively. According to results of XRD analysis (not shown) the $Cr_2AlC$ sample had about 3 % of $Cr_7C_3$ and 2 % of $Al_2O_3$ as impurities. In the $Ti_2AlC$ sample the main impurities were $Ti_3AlC_2$ (2 %) and TiC (2 %). Each couple is discussed separately below.

### 3.2 $Cr_2AlC$/NSA Diffusion Couple

SEM micrographs of the bonded interface of $Cr_2AlC$/NSA diffusion couples are presented in Figure 1. No evidence of cracking and/or delamination at the interface were observed. A complex, multiphase diffusion zone, DZ, however, did form. The overall thickness of the DZ (Figure 1) was of about 18 µm after HP at 1000°C (Figure 1a) and 30 µm after treatment at 1100°C (Figures 1b and 1c). There was no difference in thickness and morphology of the DZ when the applied pressure was increased from 2 to 20 MPa. Unaffected $Cr_2AlC$ and NSA structures can be seen at the extreme left and right regions of the SEM images, respectively. The presence of small dark regions in the reaction layer, that appear to be similar in shape and morphology to those in $Cr_2AlC$ suggest that the location of the original interface can be given by the yellow line in Figure 1a.

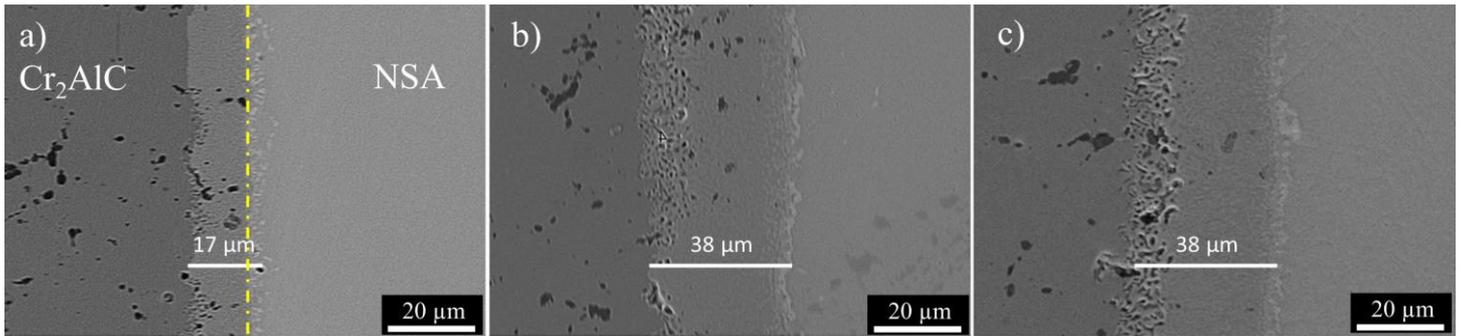

**Figure 1.** SEM micrographs of the Cr$_2$AlC/NSA interface after HP for 2 h at, a) 1000°C, b) and, c) 1100°C; in a) and b) the load corresponded to a stress of 2 MPa; in c) the stress was 20 MPa. Yellow line in (a) denotes location of original interface.

The microstructure and composition of the interface region after treatment at 1100°C under 20 MPa are shown in Figures 2a and 2b, respectively. Three main regions in the DZ are observed. The exhibited EDS spectra (Figure 2b) suggest that Region I is a Cr$_7$C$_3$ layer. Region II is most probably an intimate mixture of Cr$_7$C$_3$ and β-NiAl, with an increased amount of Cr$_7$C$_3$ observed closer to Region I. Region II clearly shows the migration of Ni, and the depletion of Al on the Cr$_2$AlC side. Finally, Region III is mainly composed of α-Cr(Mo) phase due to the out diffusion of Ni from that region.[43] In general, the obtained microstructure shows many similarities to prior detailed studies of Cr$_2$AlC/LSHR diffusion couples that also show three main regions in the DZ. [14]

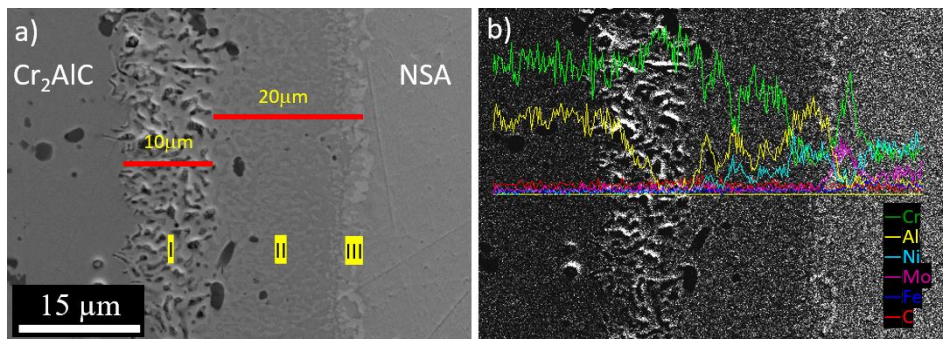

**Figure 2.** a) SEM micrographs of the Cr$_2$AlC/NSA interface after HP at 1100°C for 2 h under 2 MPa; b) Corresponding elemental EDS line scan.

## 3.3 Ti$_2$AlC/NSA Diffusion Couple

In contrast to the Cr$_2$AlC/NSA diffusion couple, there was no bonding between the Ti$_2$AlC and the NSA after treatment at 1000°C when the stress was 2 MPa. When the temperature was increased to 1100°C, under the same pressure, partial bonding was achieved (Figure 3a). However, it was only when the stress was increased to 20 MPa that full bonding, without cracks or delamination, was realized (Figures 3b and 3c). The overall thickness of the DZ was around 60 µm.

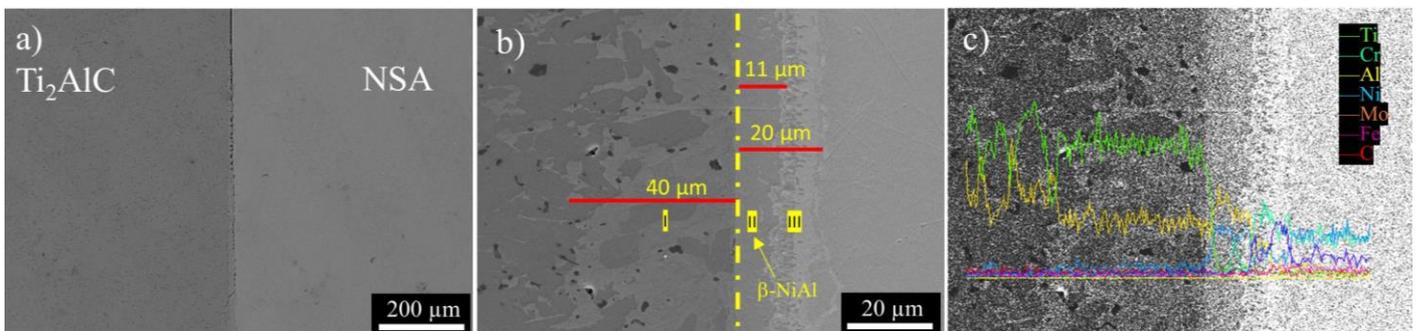

**Figure 3.** SEM micrographs of the Ti2AlC/NSA interface after HP at, a) 1100°C and a load corresponding to a stress of 2 MPa, b) 1100°C and stress of 20 MPa and, c) corresponding elemental EDS line scan of (b). Yellow line in (b) denotes location of original interface.

Poor surface preparation, characterized by residual scratches and roughness, can be one of the reasons for the poor bonding between Ti$_2$AlC and NSA.[44] We note in passing that both Ti$_2$AlC and Cr$_2$AlC discs were prepared in the same exact way and several HP bonding attempts were made with different Ti$_2$AlC/NSA samples at lower stresses and temperatures, but none were successful. A more likely reason for the poor bonding could be the relatively large CTE differences, that can lead to thermal stress between Ti$_2$AlC (~8·10$^{-6}$ °C$^{-1}$) [36,37] and NSA (≥16·10$^{-6}$ °C$^{-1}$) upon cooling.[39]

Nevertheless, at the higher stress, the contact area between the Ti$_2$AlC and the NSA surface is larger and hence the formation kinetic of initial DZ is enhanced.[45]

Based on XRD (Figure 4) and EDS point analysis (not shown) this diffusion layer (marked as Region II on Figure 3b) corresponds to β-NiAl. Fortunately, the CTE of this newly formed β-NiAl layer fits in between the CTE values of Ti$_2$AlC and NSA, [45] that in turn leads to a better match and to lower thermal stresses. In addition to the β-NiAl layer, two additional regions are discernible: a 40 µm thick (marked as Region I on Figure 3b) on the Ti$_2$AlC side of the interface and a 10 µm thick on the NSA side (marked as Region III on Figure 3b).

From the XRD pattern (Figure 4) combined with EDS point analysis (not shown) of Region I (Fig. 4b), the newly formed phase on the MAX side corresponds to Ti$_3$NiAl$_2$C. It is important to point here that several attempts were made to obtain quantitative information from XRD patterns of the interfaces of the various diffusion couples, but no useful information was obtainable, mainly because of the fracture roughness and the small volume of the diffusion layer. Said otherwise, because of the poor quality of the XRD patterns. Moreover, if the cleavage doesn't pass exactly where the reaction layer was, the possibility of obtaining a decent XRD signal is low. Finally, Region III was similar in composition and morphology to the phase observed in Region III in of Cr$_2$AlC case

In comparison to the Cr$_2$AlC/NSA case, for the Ti$_2$AlC/NAS diffusion couple there is no ambiguity as to the location of the original interface since the outline of the Ti$_2$AlC grains are still clearly visible to the left of the thin yellow line shown in Figure 3b. Based on the EDS analysis, the DZ constituents can be described in terms of NSA elements (mainly Ni and Cr) diffusing into the MAX and the counter diffusion of the A element (Al) into the NSA. The diffusion of M and X elements into the NSA was not observed.

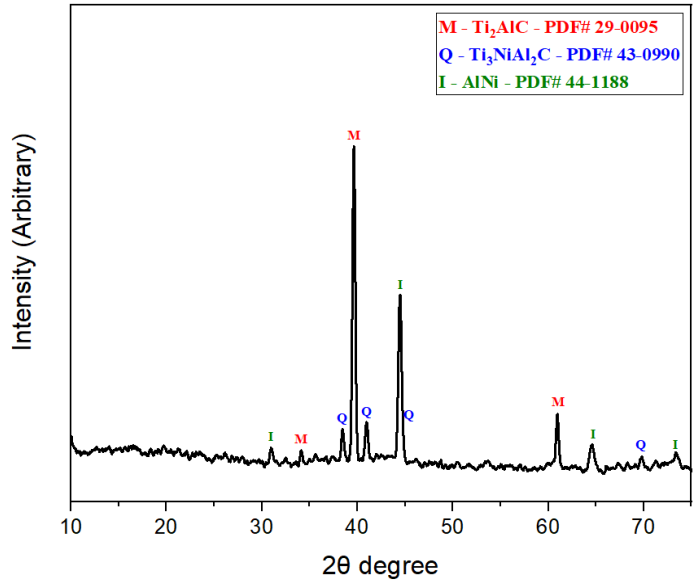

**Figure 4.** XRD patterns of the interface at the Ti$_2$AlC side of Ti$_2$AlC/ Inc718 diffusion couple after heating at 1100°C under 20 MPa.

## 3.4 Thermal Cycling

A collage made from SEM micrographs of the oxidized post thermal cycled interface of the Ti$_2$AlC/NSA/Cr$_2$AlC diffusion couple is shown in Figure 5a. While delamination cracks and oxidation of the Ti$_2$AlC/NSA interface was observed after ≈ 8 cycles, the Cr$_2$AlC/NSA interface showed no signs of delamination nor oxidation even after 20 cycles.

Higher magnification of the interfaces after thermal cycling of the Ti$_2$AlC/NSA and Cr$_2$AlC/NSA are presented in Figures 5b and 5c respectively. The cycling did not result in significant changes in either the morphology or composition of the various regions of the DZ for the Cr$_2$AlC/NSA couple. Although, regions I and II grew from 10 and 20 to 20 and 30 μm, respectively (Figure 5c). Additionally, many small black particles with spherical morphology appeared at the bonding interface between Cr$_2$AlC and NSA (marked by two red dashed lines in Figure 5c). Based on EDS point analysis, these black dots correspond to $Al_2O_3$. On the $Ti_2AlC$/NSA side, both $Ti_2AlC$ and Ni were oxidized on both edges of the cracks. While the oxide layer on the $Ti_2AlC$ side corresponds to $Al_2O_3$, on the NSA side the oxide layer consists mainly of $Cr_2O_3$ with small amounts of $Al_2O_3$.

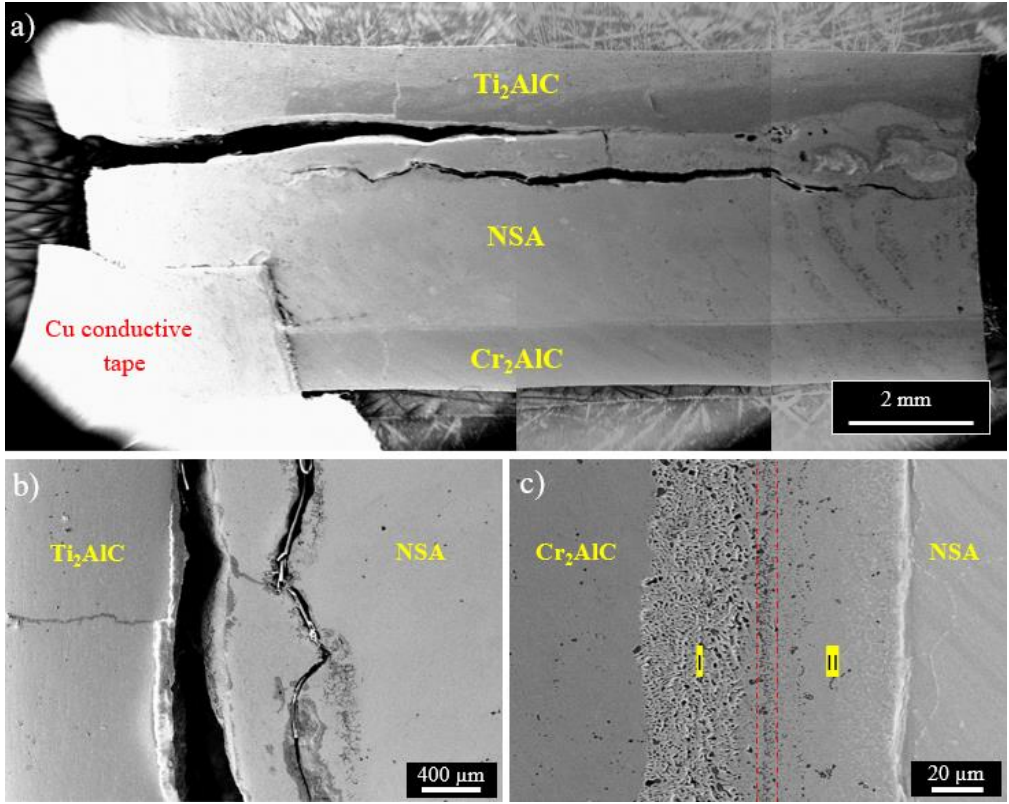

**Figure 5.** a) Collage of cross-sectional SEM micrographs of the Ti2AlC/NSA/Cr2AlC diffusion couple taken after 20 thermal cycles, b) higher magnification of $Ti_2AlC$/NSA interface and, c) higher magnification of $Cr_2AlC$/NSA interface.

## 4. Summary and Conclusions

Herein we report, for the first time, on the interaction of Inconel 718 with two different alumina-forming MAX phases, $Cr_2AlC$ and $Ti_2AlC$. Successful diffusion bonding was realized with no evidence of interfacial damage or cracking after cooling to room temperature in both cases. Adequate bonding between $Cr_2AlC$ and NSA was achieved after HP treatment at 1000°C and very low contact pressure of only 2 MPa, while in the case of $Ti_2AlC$ a higher temperature (1100°C) and considerably higher pressure of 20 MPa was needed.

In both cases the diffusion zone composition reflects constituents can be described in terms of diffusion of NSA elements (mainly Ni and Cr) into the MAX phase and counter diffusion of the A element (Al) into the superalloy, resulting in the formation of β-NiAl and new carbides. The diffusion of M and X elements into the NSA was not observed. Many of the features in the diffusion zone are reminiscent of the microstructures formed in previous studies on MAX/superalloys[14,16] and aluminized superalloys.[46,47] Likewise, in the present study, some characteristics of the interface zone were manifested on both sides of the interface as formation of new carbide ($Cr_7C_3$ and $Ti_3NiAl_2C$) and a β-NiAl phase.

Thermal cycling experiments showed that $Ti_2AlC$ is a poor candidate as an oxidation barrier for NSA at temperatures in the vicinity of 1000°C. However, in the case of $Cr_2AlC$/NSA there was no cracks, delamination nor surface degradation. Previous studies showed that $Cr_2AlC$ have extremely slow oxidation rate at elevated temperatures.[16,31] Combined together these results make the $Cr_2AlC$ MAX as a promising candidate as a high temperature, corrosion resistant and strain tolerant diffusion barrier for Ni-based superalloys.

## 5. Acknowledgements

The authors acknowledge funding from the Euratom research and training programme 2014-2018 under grant agreement No. 740415.